\documentclass[preprint,5p,times,twocolumn]{elsarticle}

\biboptions{sort&compress}

\usepackage{amsmath,amsfonts}
\usepackage{algorithmic}
\usepackage{algorithm}
\usepackage{array}
\usepackage[caption=false,font=normalsize,labelfont=sf,textfont=sf]{subfig}
\usepackage{textcomp}
\usepackage{stfloats}
\usepackage{url}
\usepackage{verbatim}
\usepackage{graphicx}
\usepackage{cite}

\usepackage{package}
\usepackage{xcolor}
\newcommand{\revised}[1]{{\color{black}#1}}

\usepackage{amssymb}

\journal{Information and Software Technology}

\begin{document}

\begin{frontmatter}



\title{Exploring Individual Factors in the Adoption of LLMs\\for Specific Software Engineering \revised{Purposes}}


\author[label1]{Stefano Lambiase} 
\author[label2]{Gemma Catolino} 
\author[label2]{Fabio Palomba} 
\author[label2]{Filomena Ferrucci} 
\author[label1]{Daniel Russo} 

\affiliation[label1]{organization={Aalborg University Copenhagen},
            addressline={A. C. Meyers Vænge, 15}, 
            city={Copenhagen},
            postcode={2450},
            country={Denmark}}

\affiliation[label2]{organization={University of Salerno},
            addressline={Via Giovanni Paolo II, 132}, 
            city={Fisciano (SA)},
            postcode={84084}, 
            country={Italia}}

\begin{abstract}
\textbf{Context:} The advent of Large Language Models (LLMs) is transforming software development, significantly enhancing software engineering (SE) processes. Research has explored their role within development teams, focusing on the specific \revised{purposes} for which LLMs are used within SE tasks, such as artifact generation, decision-making support, and information retrieval. Despite the growing body of work on LLMs in SE, most studies have centered on broad adoption trends, neglecting the nuanced relationship between individual cognitive and behavioral factors and their impact on \revised{purpose-specific} adoption. While factors such as perceived effort and performance expectancy have been explored at a general level, their influence on distinct SE \revised{purposes} remains underexamined. This gap hinders the development of tailored LLM-based systems (e.g., Generative AI Agents) that align with engineers' specific needs and limits the ability of team leaders to devise effective strategies for fostering LLM adoption in targeted workflows. 
\textbf{Objectives:} For the reasons mentioned above, this study aims to study the individual factors that drive the choice to use LLMs for distinct SE \revised{purposes}.
\textbf{Methods:} To achieve the above-mentioned objective, we surveyed 188 software engineers to test the relationship between individual attributes related to technology adoption and LLM adoption across five key \revised{purposes}, using structural equation modeling (SEM). The Unified Theory of Acceptance and Use of Technology (UTAUT2) was applied to characterize individual adoption behaviors. 
\textbf{Results:} The findings reveal that \revised{purpose-specific} adoption is influenced by distinct factors, some of which negatively impact adoption when considered in isolation, underscoring the complexity of LLM integration in SE.
\textbf{Conclusions:} To support effective adoption, this article provides actionable recommendations, such as seamlessly integrating LLMs into existing development environments and encouraging peer-driven knowledge sharing to enhance information retrieval.
\end{abstract}


\begin{highlights}
\item Performance Expectancy strongly drives both intention and AI-assisted use.
\item Habit is the strongest predictor across all types of AI-supported behaviors.
\item Facilitating Conditions shape actual use but do not affect adoption intention.
\item AI’s impact differs across \revised{purposes}, from decision support to training activities.
\item Social Influence plays a limited role, except in \revised{purposes} requiring shared context.
\end{highlights}

\begin{keyword}
Software Engineering \sep Management \sep Technology Adoption \sep PLS-SEM \sep Generative AI \sep LLM



\end{keyword}

\end{frontmatter}

\section{Introduction}

The rise of Generative AI (GenAI), particularly its most prominent incarnation—Large Language Models (LLMs)—has profoundly transformed various domains, earning them the label of disruptive technologies. Among these, software development has been significantly impacted, as LLMs continue to reshape how stakeholders engage with this activity. Several studies already highlight how LLMs have enhanced developers' workflows, with many reporting a perceived increase in productivity\revised{~\citep{liang2024large}}. Indeed, LLMs can support developers in various ways, for example, by automating code generation~\citep{liu2024empirical, Bhattacharya2023Exploring, Jiang2023Self_planning} and—more in general—assisting in everyday activities~\citep{Dong2023Self_collaboration, yao2024survey, ross2023programmer}.

Given its potential, research in software engineering (SE) has sought to characterize how professionals utilize these tools for various development \revised{purposes}. Indeed, effectively managing the integration of these tools requires first understanding the specific \revised{purposes} for which they are adopted---an understanding that depends on systematic measurement. As expected, these \revised{purposes} range from relatively simple to highly diverse. LLMs are widely reported to be used for source code generation~\citep{liu2024empirical, Bhattacharya2023Exploring, Jiang2023Self_planning} and comprehension support~\citep{Bhattacharya2023Exploring,kumar2024code}, as well as for assisting in software testing through the generation of test cases and test suites~\citep{Schäfer2023An, Tang2023ChatGPT, Wang2023Software}. Additionally, they are employed, more generally, as decision-support assistants for developers~\citep{Dong2023Self_collaboration, yao2024survey, ross2023programmer}. Moreover, a more recent study by Khojah et al.~\citep{khojah2024beyond} proposed a higher-level theoretical framework regarding how software engineers utilize LLMs, specifically ChatGPT. Analyzing their work and related literature, it is possible to identify five broad categories of use: \textit{artifact manipulation}, \textit{generation of alternative versions of the same artifact}, \textit{expert consultation} (intended as decision support), \textit{information retrieval}, and \textit{training} (understood as engaging in conversation to learn new concepts).

As emerging from the existing literature, an initial understanding of the \revised{purposes} most frequently pursued in collaboration with LLMs has emerged in the context of software engineering. However, what remains insufficiently explored is the antecedent of use and the underlying motivation driving the adoption of LLMs for specific \revised{purposes}. Several studies have investigated individual factors related to technology adoption—e.g., perceived effort and performance expectancy—influencing developers' decisions to use LLMs in software engineering~\citep{russo2024_navigating, lambiase2024investigatingroleculturalvalues, choudhuri2024guideschoicesmodelingdevelopers}. 
For example, in our previous work~\citep{lambiase2024investigatingroleculturalvalues}, we investigated how individual developers' cultural values—based on Hofstede's model—affect the general intention to adopt LLMs in software development, using the UTAUT2 model as a mediating framework. That study focused on technology acceptance at a global level, regardless of purpose type. 
Nevertheless, these efforts have primarily focused on overall adoption, without examining \textbf{the individual factors that drive the choice to use LLMs for distinct software engineering \revised{purposes}}. 

We argue that the highlighted gap is problematic for several reasons. First, current research in software engineering is focused on developing tools that incorporate LLMs to support specific \revised{purposes}~\citep{white2024building, nguyen2023generative}; a focused investigation into which individual developer factors are most strongly associated with a higher frequency of use for specific \revised{purposes} could inform design choices, support developer decision-making, and enhance the development of tailored tools, ultimately reducing wasted effort. Second, gaining insight into these motivations is essential for organizations aiming to optimize the integration of LLMs into their software development processes, ensuring that their use aligns with productivity objectives while mitigating potential unintended consequences. Finally, exploring these antecedents could inform studies on long-term behavioral changes, shedding light on how reliance on LLMs evolves over time and its implications for developer skill acquisition, software quality, and team dynamics.


The objective of this work is to address the above-mentioned research gap by outlining a developer profile, defined by individual factors related to technology adoption, that characterizes the choice of using an LLM for specific SE \revised{purposes}. Specifically, this study aims to address the following research question.

\steResearchQuestionBox{\faQuestionCircle\ \textbf{RQ}—\textit{What individual factors influence developers' choices to use LLMs for specific SE \revised{purposes}?}}

To do so, we built upon the dataset collected for our previous study~\citep{lambiase2024investigatingroleculturalvalues} by extending it with additional \revised{purpose-specific} adoption measures and shifting the analytical focus, investigating how UTAUT2 constructs influence the actual frequency of LLM use across distinct software engineering \revised{purposes}. Specifically, we used a validated questionnaire, surveying $N=188$ software engineers. Data has been subsequently analyzed with Partial Least Squares Structural Equation Modeling (PLS-SEM). To operationalize the dependent variables—i.e., the adoption of LLMs for software engineering \revised{purposes}—the framework proposed by Khojah et al.~\citep{khojah2024beyond} was adopted, as it provides a comprehensive and generalizable representation of LLM adoption in software engineering. For the independent variables—i.e., developer profiles—we used the UTAUT2 framework~\citep{Venkatesh_2012_UTAUT2}, given its robustness and reliability in capturing the key factors influencing technology adoption. 

The results depict a heterogeneous adoption landscape. Different developer profiles appear to favor distinct \revised{purposes}, with certain characteristics even counterintuitively disadvantaging specific uses. For instance, the extent to which a developer values peer opinions is associated with a higher likelihood of using LLMs for \revised{purposes} related to retrieving information for decision-making. Conversely, strong support for LLM adoption appears to discourage their use for the same purposes, suggesting a nuanced and complex integration process.

\section{Related Work}

LLMs have demonstrated considerable potential across various domains, including software engineering~\citep{brown2020language, khojah2024beyond, barke2023grounded, della2025prompt, dellaporta2026prompt, dellaporta2026rebuilding}. Within this specific context, LLMs are increasingly being employed for a wide range of purposes, extending beyond code generation~\citep{liu2024empirical, Jiang2023Self_planning} to include tasks such as code summarization, explanation~\citep{Bhattacharya2023Exploring,kumar2024code}, and general developer support~\citep{Dong2023Self_collaboration, yao2024survey, ross2023programmer}. Moreover, beyond examining the motivations for adoption, recent software engineering research has also focused on analyzing the adoption of LLMs itself, investigating the factors that influence their integration into development workflows. The remainder of this section presents an overview of the studies that are most closely related to this work.

\subsection{Use of Large Language Models}\label{sec_LLM_task}

Barke et al.~\citep{barke2023grounded} conducted a grounded theory study with 20 participants to examine how developers interact with Copilot for source code generation. Their study identified a bimodal interaction model, distinguishing two primary modes of engagement: acceleration and exploration. In the acceleration mode, developers utilize Copilot to expedite tasks when they have a clear plan of action. Conversely, the exploration mode is adopted when developers face uncertainty, using Copilot to investigate and generate potential solutions.

Similarly to the previous work, Khojah et al.~\citep{khojah2024beyond} conducted an empirical study with 24 engineers, developing a theoretical framework that describes the adoption of LLMs for specific SE \revised{purposes}. This framework identifies five key purposes: artifact generation and modification, alternatives generation, decision support, information retrieval, and training (engaging in conversational learning to acquire new concepts). Additionally, the study examines how these purposes relate to both internal and external factors stemming from LLM usage, as well as their impact on personal experiences in terms of trust and perceived usefulness.

\revised{Focusing more specifically on code generation, LLMs have shown promising results in this domain, although their effectiveness remains context-dependent and an active area of investigation. When appropriately guided through prompt engineering techniques, these models can generate code that meets practical software engineering requirements in several reported scenarios.} This capability has been empirically demonstrated by Liu et al.~\citep{liu2024empirical} and Jiang et al.~\citep{Jiang2023Self_planning}. Liu et al.~\citep{liu2024empirical} introduced the Prompt-FDC method, which improves code completeness, enhances comment clarity, and supports the generation of code for safety-critical applications. Jiang et al.~\citep{Jiang2023Self_planning} proposed a structured self-planning framework for code generation, incorporating distinct planning and implementation phases to improve the overall quality and coherence of the generated code.

Within the domain of code generation, LLMs are increasingly being explored for their potential in generating test cases and test suites. Specifically, in the context of unit testing, Schäfer et al.~\citep{Schäfer2023An} and Tang et al.~\citep{Tang2023ChatGPT} have proposed notable approaches. Schäfer et al.~\citep{Schäfer2023An} leveraged prompt engineering techniques to instruct LLMs in generating unit tests, achieving results comparable to human-written test code in terms of both coverage and quality. Tang et al.~\citep{Tang2023ChatGPT} conducted a comparative analysis between LLM-generated test code and tests produced by EvoSuite, highlighting both the strengths and limitations of LLMs in this context. Moreover, beyond unit testing, Wang et al.~\citep{Wang2023Software} conducted a systematic literature review to examine the broader application of LLMs in software testing. Their study identified key challenges and opportunities associated with LLM adoption in testing activities, providing insights that shape future research directions in this area.

\revised{LLMs have also shown encouraging results in the areas of code summarization and program comprehension.} Kumar and Chimalakonda~\citep{kumar2024code} successfully leveraged federated learning (FedLLM) for this purpose, highlighting the advantages of distributed training methods in optimizing LLM performance. Bhattacharya et al.~\citep{Bhattacharya2023Exploring} conducted an evaluation of various LLMs in generating natural language summaries for code snippets. Their findings revealed that models fine-tuned specifically for code-related tasks consistently outperformed general-purpose models, emphasizing the importance of domain-specific adaptation for enhanced performance.  

More broadly, LLMs are increasingly utilized as software developer assistants for a wide range of programming-related tasks. Ross et al.~\citep{ross2023programmer} conducted experiments with human participants to assess LLMs as programming assistants. Although participants initially expressed skepticism, they were ultimately impressed by the assistants' extensive capabilities, the quality of their responses, and their potential to enhance productivity. \revised{Furthermore, LLMs have shown potential in strengthening code security, including vulnerability detection and data privacy protection, with some studies reporting improvements over traditional methods in specific settings.} However, as Yao et al.~\citep{yao2024survey} caution, the advanced reasoning capabilities of LLMs also introduce risks, such as potential misuse for user-targeted attacks.  

An additional noteworthy contribution in the domain of collaborative assistants was made by Dong et al.~\citep{Dong2023Self_collaboration}. Their work explored the creation of a virtual team composed of LLM-powered agents, designed to autonomously address various software engineering and development tasks. By integrating established software engineering methodologies into the workflow, \revised{the authors reported that these agents could carry out software development tasks with a degree of autonomy, suggesting potential for reducing the need for direct human intervention in certain scenarios.}

\subsection{Adoption of Large Language Models}

The research community in SE has already started to study the factors influencing the adoption of LLM.

Two of the most important factors appear to be the perceived usefulness and frequency of use of LLMs, as users are more inclined to integrate LLMs into their workflows when they perceive them as valuable for their tasks~\citep{agossah2023llm, khojah2024beyond}. Agossah et al.~\citep{agossah2023llm} conducted a survey study among IT company employees to examine their intention to adopt generative AI tools, revealing that perceived usefulness plays a crucial role in influencing adoption decisions. Their findings also highlight the importance of frequency of use in shaping employee perceptions, suggesting the need for further research into how different user groups engage with these tools.

Habit has also been identified as a fundamental factor in LLM adoption. Draxler et al.~\citep{draxler2023gender} analyzed LLM usage patterns among 1500 U.S. citizens, uncovering a gender disparity in adoption and emphasizing the role of technology-related education in mitigating this gap. Their findings indicate that experienced users are more likely to utilize LLMs for professional tasks, underscoring the importance of equitable AI education and access to maximize workplace benefits.

Another critical aspect is the ease of integration of LLMs into existing workflows and their overall usability. Russo~\citep{russo2024_navigating} investigated the adoption of generative AI in software engineering, introducing the Human-AI Collaboration and Adaptation Framework (HACAF), which synthesizes multiple theoretical perspectives to provide a comprehensive understanding of adoption dynamics. This mixed-method study involved a qualitative investigation to develop the framework, followed by a PLS-SEM statistical analysis to validate it. The findings indicate that social influence is not a significant factor, whereas the degree to which the tool seamlessly integrates into professional activities is the primary determinant of adoption.

 Similarly to the Russo’s work~\citep{russo2024_navigating}, a previous study conducted by the authors~\citep{lambiase2024investigatingroleculturalvalues} employed the UTAUT2 framework alongside Hofstede’s cultural dimensions to examine the role of cultural values in the adoption of LLMs for software engineering \revised{purposes}. The findings indicated that cultural values do not appear to significantly influence LLM adoption. Instead, habitual usage emerged as the primary predictor of both the intention to use and the actual adoption of these models. 
\footnote{In the present work, we shift the analytical focus: rather than modeling general adoption through cultural values, we investigate how UTAUT2 constructs relate to the frequency of LLM usage across specific SE purposes. This means that we both changed the dependent and independent variable.}
Additionally, Choudhuri et al.~\citep{choudhuri2024guideschoicesmodelingdevelopers} recently explored the role of trust in LLM adoption, focusing on developers within two large firms. Their study revealed that trust is closely linked to actual usage, with its influence shaped by specific antecedents.

\revised{From a complementary, large-scale behavioral perspective, Brown et al.~\citep{brown2024trust} analyzed approximately one million multi-line code completion suggestions seen by \num{59000} developers at Google, using acceptance rate as a proxy for trust. Their mixed-methods study identified suggestion quality, developer language expertise, and development context as key factors influencing whether developers accept AI-generated code completions, highlighting the multi-faceted nature of trust in AI-powered development tools.} 

\steSummaryBox{\faList \hspace{0.05cm} Related Work: Summary and Research Gap.}{The research community is dedicating significant effort to characterize the role of LLMs in software engineering. However, despite the growing body of research examining the factors influencing LLM adoption, there remains a gap in understanding the individual factors associated with increased usage of LLMs for specific purposes. Expanding the current body of knowledge with insight into these individual determinants could complement the Khojah et al.~\citep{khojah2024beyond} framework and provide valuable guidance for development teams, allowing them to better direct their efforts in supporting developers' effective use of LLMs.}

\section{Objective and Research model}\label{sec_hyp}

To address the identified research gap, this study aims to investigate whether specific individual factors, which characterize the profile of professionals involved in software development, are associated with a higher or lower use of LLM for specific software engineering purposes. To achieve this objective, we identified (1) a comprehensive representation of the software engineering purposes most frequently pursued by professionals using LLMs (dependent variables) and (2) a detailed representation of the factors influencing an individual's decision to adopt this technology (independent variables).

\begin{figure}
    \centering
    \includegraphics[width=1\linewidth]{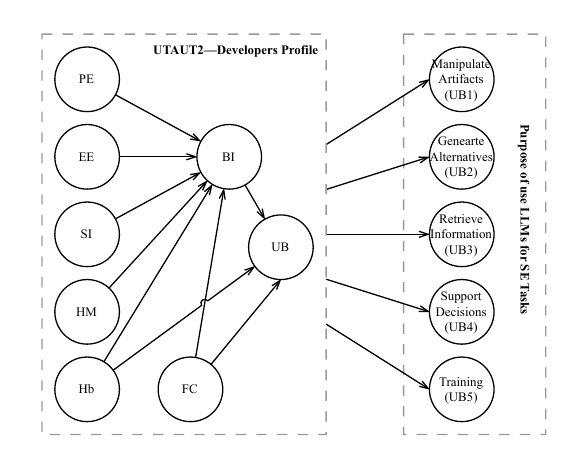}
    \caption{Research model and hypotheses.}
    \label{fig_research_model}
\end{figure}

\subsection{\revised{Characterizing Purposes for Adopting LLMs in SE}}\label{sec_purpose_of_adoption}

\revised{As discussed in Section \ref{sec_LLM_task}, there are few frameworks for categorizing the purposes for which LLMs are adopted in software engineering. Given our broad and general perspective, the taxonomy proposed by Khojah et al.~\citep{khojah2024beyond} was used as the foundation for operationalizing our dependent variables. Khojah et al.'s taxonomy identifies three high-level purposes---artifact manipulation, expert consultation, and training---each further decomposed into subtypes: artifact manipulation includes artifact generation, artifact modification, brainstorming, and side tasks; expert consultation includes problem solving, information retrieval, and decision making; and training includes drill-down learning and learning by example. Since our study required a level of granularity suitable for survey-based operationalization, we consolidated certain subtypes while preserving conceptually meaningful distinctions. The resulting five categories are described below.}


\begin{itemize}
    \item \textit{Manipulate Artifacts (UB1)}—Utilizing LLMs to generate new software artifacts or modify one, including source code, documentation, and other software-related deliverables.
    \item \textit{Generate Alternatives (UB2)}—Leveraging LLMs to produce alternative versions of an artifact that differ based on specific attributes.
    \item \textit{Information Retrieval (UB3)}—Using LLMs to extract and synthesize information from external sources, such as forums and technical documentation.
    \item \textit{Decision Support (UB4)}—Engaging LLMs in conversations aimed at facilitating decision-making through recommendations and contextual insights.
    \item \textit{Training (UB5)}—Employing LLMs to acquire theoretical or practical knowledge on software concepts, often requiring multi-step interactions.
\end{itemize}

A variable was included in the research model for each of the aforementioned purposes, as shown on the right side of Figure \ref{fig_research_model}. Each variable indicates the frequency of use of LLMs for a specific purpose.

\subsection{Characterizing Developer Profiles in LLM Adoption}

For the independent variables, developer profiles were represented using the constructs of the UTAUT2 model~\citep{Venkatesh_2012_UTAUT2}, a widely recognized framework for explaining technology adoption. UTAUT2 has been applied in multiple studies within the software engineering domain~\citep{russo2024_navigating, lambiase2024investigatingroleculturalvalues}, making it a suitable foundation for characterizing LLM adoption patterns across specific software engineering \revised{purposes}.

Venkatesh et al.~\citep{Venkatesh_2003_UTAUT} developed the original Unified Theory of Acceptance and Use of Technology (UTAUT) by synthesizing eight prior models, identifying four key constructs that predict behavioral intention (BI) and usage behavior (UB): 
(1) \textit{Performance Expectancy} (PE), referring to the belief that a system enhances job performance; 
(2) \textit{Effort Expectancy} (EE), reflecting the perceived ease of use; 
(3) \textit{Social Influence} (SI), representing the degree to which users perceive that important others endorse system adoption; and 
(4) \textit{Facilitating Conditions} (FC), referring to the perceived availability of technical and organizational support. 
Subsequently, UTAUT was extended to UTAUT2~\citep{Venkatesh_2012_UTAUT2} to better account for technology adoption in consumer settings, introducing three additional constructs: 
\textit{Hedonic Motivation} (HM), which captures the enjoyment derived from using a system; 
\textit{Price Value} (PV), reflecting users' cost-benefit evaluation of a technology; and 
\textit{Habit} (Hb), representing the extent to which individuals engage in automatic behaviors based on learning and experience. 

From UTAUT2, seven constructs were included in this study: performance expectancy (PE), effort expectancy (EE), social influence (SI), hedonic motivation (HM), facilitating conditions (FC), habit (Hb), behavioral intention (BI), and usage behavior (UB). The price value (PV) was omitted, given that many developers in our sample utilize LLM-based tools without direct financial costs. All included constructs have been shown to positively influence technology adoption, as they capture different aspects of how favorably users perceive the technology~\citep{Venkatesh_2012_UTAUT2, russo2024_navigating, lambiase2024investigatingroleculturalvalues}. Moreover, behavioral intention serves as a key mediator, linking individual perceptions of LLMs to actual usage~\citep{Venkatesh_2012_UTAUT2, lambiase2024investigatingroleculturalvalues}. Additionally, both habit and facilitating conditions have been found to directly influence actual usage, independent of behavioral intention~\citep{Venkatesh_2012_UTAUT2, lambiase2024investigatingroleculturalvalues}.

\subsection{The Resulting Research Model}

The constructs mentioned earlier were tailored to fit the context of LLM adoption, providing the foundation for our research hypotheses and model in Figure \ref{fig_research_model}. Each construct is hypothesized to align with the usage purposes detailed in Section \ref{sec_purpose_of_adoption}.

\begin{itemize}
    \item[H1] \revised{The frequency of using LLMs for} manipulating artifacts (UB1) is positively impacted by:
    \begin{itemize}
        \item[H1a] performance expectancy (PE),
        \item[H1b] effort expectancy (EE),
        \item[H1c] social influence (SI),
        \item[H1d] hedonic motivation (HM),
        \item[H1e] habit (Hb),
        \item[H1f] facilitating conditions (FC),
        \item[H1g] behavioral intention (BI), and
        \item[H1h] usage behavior (UB).
    \end{itemize}
    
    \item[H2] \revised{The frequency of using LLMs for} generating alternatives of artifacts (UB2) is positively impacted by:
    \begin{itemize}
        \item[H2a] performance expectancy (PE),
        \item[H2b] effort expectancy (EE),
        \item[H2c] social influence (SI),
        \item[H2d] hedonic motivation (HM),
        \item[H2e] habit (Hb),
        \item[H2f] facilitating conditions (FC),
        \item[H2g] behavioral intention (BI), and
        \item[H2h] usage behavior (UB).
    \end{itemize}
    
    \item[H3] \revised{The frequency of using LLMs for} information retrieval (UB3) is positively impacted by:
    \begin{itemize}
        \item[H3a] performance expectancy (PE),
        \item[H3b] effort expectancy (EE),
        \item[H3c] social influence (SI),
        \item[H3d] hedonic motivation (HM),
        \item[H3e] habit (Hb),
        \item[H3f] facilitating conditions (FC),
        \item[H3g] behavioral intention (BI), and
        \item[H3h] usage behavior (UB).
    \end{itemize}
    
    \item[H4] \revised{The frequency of using LLMs for} decision-making support (UB4) is positively impacted by:
    \begin{itemize}
        \item[H4a] performance expectancy (PE),
        \item[H4b] effort expectancy (EE),
        \item[H4c] social influence (SI),
        \item[H4d] hedonic motivation (HM),
        \item[H4e] habit (Hb),
        \item[H4f] facilitating conditions (FC),
        \item[H4g] behavioral intention (BI), and
        \item[H4h] usage behavior (UB).
    \end{itemize}
    
    \item[H5] \revised{The frequency of using LLMs for} training facilitation (UB5) is positively impacted by:
    \begin{itemize}
        \item[H5a] performance expectancy (PE),
        \item[H5b] effort expectancy (EE),
        \item[H5c] social influence (SI),
        \item[H5d] hedonic motivation (HM),
        \item[H5e] habit (Hb),
        \item[H5f] facilitating conditions (FC),
        \item[H5g] behavioral intention (BI), and
        \item[H5h] usage behavior (UB).
    \end{itemize}
\end{itemize}

A key aspect of the proposed model is the mediating role of behavioral intention and usage behavior in shaping the relationships between the foundational constructs of UTAUT2 and the specific purposes for which LLMs are used in SE. BI serves as an intermediary linking individual perceptions to the general actual use of LLMs in SE (UB) and the specific adoption patterns (UB1 to UB5). Moreover, UB further mediates these relationships by translating BI, FC, and Hb into concrete usage behavior across different SE purposes (UB1 to UB5). This dual mediation pathway, defined in the original UTAUT2 model, suggests that while foundational constructs drive the inclination to adopt LLMs, their ultimate impact on purpose-specific usage is contingent on both the intention to use LLMs and the frequency with which they are integrated into developers' workflows.

\revised{Two modeling choices deserve explicit justification. First, UB was not modeled as a higher-order construct with UB1--UB5 as lower-order constructs. Although general LLM usage is a precondition for purpose-specific usage, the five purposes are not reflective indicators of a single latent construct: they represent distinct behavioral outcomes potentially driven by different factors. A higher-order specification would collapse the very variation our research question aims to investigate. Instead, UB was retained as a mediator, allowing us to examine how general usage translates into purpose-level adoption.}

\revised{Second, direct paths from PE, EE, SI, and HM to UB1--UB5 were included alongside the standard UTAUT2 mediation through BI and UB. While the original UTAUT2 does not specify such direct paths, our model extends beyond general usage behavior to purpose-specific outcomes, and there is theoretical reason to expect that individual factors may directly shape whether developers use LLMs for specific purposes---independently of general intention. For instance, social influence may directly encourage the use of LLMs for information retrieval, where peer validation is particularly relevant, without necessarily affecting overall behavioral intention first. This choice is supported by prior extensions of UTAUT in domain-specific contexts: Or~\citep{or2023revisiting}, in a meta-analytic synthesis, found that effort expectancy and social influence emerged as direct predictors of usage behavior when direct paths were introduced; similarly, Du and Lv~\citep{du2024factors} found that facilitating conditions had the strongest direct impact on actual use behavior in the context of generative AI adoption for education. We acknowledge that this increases model complexity; however, the purpose-specific analysis constitutes the core contribution of this study and warrants this trade-off.\footnote{Importantly, non-significant direct paths do not introduce bias in PLS-SEM estimation, and the significant ones reveal meaningful relationships that would otherwise be masked by mediation alone.}}

\section{Research Method}\label{sec:experiment}

To test the research model introduced in Section \ref{sec_hyp}, we conducted a survey study. Subsequently we analyzed data with \textit{Partial Least Squares Structural Equation Modelling (PLS-SEM)}~\citep{hair_2014_PLS}.

PLS-SEM is a statistical method, often compared to traditional regression, used to test and analyze complex relationships and to build predictive models~\citep{hair_2014_PLS, russo2021_pls_SLR} and has been increasingly adopted in SE research~\citep{lambiase2024investigatingroleculturalvalues, russo2021_pls_SLR, russo2024_navigating, lambiase_quantifying_2026}. Unlike standard regression, however, PLS-SEM allows for the analysis of more complex structures, including multiple dependent variables. This method is ideal for analyzing \textit{latent} variables—constructs that are not directly observable—and must be measured using validated instruments. Moreover, PLS-SEM is widely used for theory testing because it incorporates a range of statistical tests designed not only to evaluate hypotheses significance but also to assess the quality and reliability of the data.

The research design involved the following steps:
\begin{enumerate}
    \item We first developed a questionnaire using validated scales from the literature to measure the constructs included in the research model described in Section \ref{sec_hyp}.
    \item Then, we administered the questionnaire to software practitioners with experience in using LLMs for SE tasks to collect responses for measuring the constructs.
    \item Last, we used PLS-SEM to analyze the data and test the hypotheses.
\end{enumerate}
The remainder of this section elaborates on the methodological steps undertaken in detail.

\subsection{Questionnaires and Data Collection}

Data collection was conducted in two phases using two distinct questionnaires. The first questionnaire aimed to measure the UTAUT2 constructs for each participant, while the second focused on assessing the frequency of LLM usage across the defined purposes. Both instruments were designed following the widely recognized survey methodology guidelines by Kitchenham and Pfleeger~\citep{kitchenham2008_PersonalOpinionSurveys}, ensuring methodological rigor in software engineering research.

As recommended by Kitchenham and Pfleeger~\citep{kitchenham2008_PersonalOpinionSurveys}, the surveys were fully anonymous, with an introductory section providing participants with all necessary details regarding the study. Each questionnaire concluded with an open-ended question for feedback and included four attention-check items to ensure response reliability. The survey was categorized within the cross-sectional survey typology, aligning with best practices for capturing perceptions at a specific point in time. To enhance validity, previously validated instruments from the literature were adopted instead of developing new measurement scales (further details on this in Section \ref{sec_measurement_instruments}).  

Guidelines by Kitchenham and Pfleeger also informed the questionnaire's design and presentation. Formatting tools available in Qualtrics were used to emphasize key elements (e.g., bold and italic fonts), and questions were systematically organized into sections to improve clarity. Additionally, pre-screening procedures were implemented to ensure that only eligible participants were included in the study. Given that the survey was conducted online, best practices outlined by Andrews et al.~\citep{andrews2007_survey_guidelines} were also followed. To adhere to these recommendations, Qualtrics was used to (1) ensure compatibility across different devices, (2) prevent multiple submissions, (3) provide an adaptive template for various browsers and screen sizes, and (4) allow participants to leave feedback.

An iterative pilot testing approach was employed to (1) evaluate the clarity and quality of the survey and (2) estimate completion time. Initially, three rounds of pilot testing were conducted with 10 researchers from our network, leading to refinements based on their feedback. Subsequently, a pilot test for each questionnaire was conducted on the data collection platform, involving five qualifying participants. \revised{The final pilot revealed no further issues. It is important to note that only the responses from the five Prolific pilot participants---who met all eligibility criteria and underwent the same filtering process as regular respondents---were included in the final dataset. The responses from the 10 researchers involved in the initial pilot rounds were used solely for questionnaire refinement and were not included in the analysis.}\footnote{Ethical approval for this study was obtained from the ethical board of Aalborg University (Case No.: 2024-505-00242).}

\subsection{Measurement Instruments}\label{sec_measurement_instruments}

The questionnaire items measuring the core constructs of the UTAUT2 model were adapted from the original work by Venkatesh et al.~\citep{Venkatesh_2012_UTAUT2}, as they have been extensively validated in the literature.\footnote{The complete list of items can be found in the appendix~\citep{online_appendix}.} The frequency of actual LLM usage for software engineering tasks (\textit{Use Behavior, UB}) was assessed using a single-item, 6-point frequency Likert scale. The seven predictor constructs were measured using a 7-point Agreement Likert scale, with the following number of items per construct: \textit{Performance Expectancy} (PE, 5 items), \textit{Effort Expectancy} (EE, 6 items), \textit{Social Influence} (SI, 5 items), \textit{Hedonic Motivation} (HM, 3 items), \textit{Facilitating Conditions} (FC, 4 items), \textit{Habit} (HB, 4 items), and \textit{Behavioral Intention} (BI, 3 items). Examples of items are “I find the use of LLM useful in supporting my job” and “People who influence my behavior think that I should use LLMs to support my job”.

Regarding the dependent variable, no validated instruments were available in the literature; therefore, it was necessary to develop a measurement approach. This process involved the entire research team, ultimately leading to the decision to adapt the validated UB item from Venkatesh et al.~\citep{Venkatesh_2012_UTAUT2}. The item was modified to incorporate the core definition of each specific usage purpose. In addition, to improve clarity, examples were provided for each purpose to ensure that respondents fully understood their meanings. For example, we developed the following item for UB1: “How frequently do you use LLMs to generate/modify artifacts (for example, code snippets, documents, etc.)”.
This strategy allowed us to rely on a well-established measurement approach while ensuring that each usage purpose was effectively communicated to participants. Consequently, the frequency of LLM usage for each purpose was measured using a single item, six-point frequency Likert scale, resulting in five additional items in the questionnaire.

In addition to the above, various demographic measures were also collected, such as gender, age, years of experience, and role. These measures allowed the recruitment phase to be balanced to ensure a heterogeneous, representative sample that is robust to influences from control variables.

\subsection{Participants Recruitment and Selection}

This section provides details on the participant recruitment process and the characteristics of the final sample.

\subsubsection{Participants Recruitment}

Participants were chosen from a group of 188 practitioners, who went through a detailed multistage recruitment process, as described by a previous study of ours~\citep{lambiase2024investigatingroleculturalvalues}. Recruitment was conducted using Prolific,\footnote{Prolific (\url{www.prolific.com}) [October 2025]} a platform designed for academic research with integrated participant filtering capabilities. Prolific is increasingly used in software engineering studies and is recognized as a high-quality recruitment platform when specific criteria—like those we adhered to as outlined—are met~\citep{eyal2021data, douglas2023data}.

In order to ensure reliability, the recruitment process adhered to the protocol established by Russo~\citep{russo2022recruiting}, incorporating the methodological choices described following.

\begin{itemize}
    \item An iterative pre-screening approach was applied to refine participant selection.
    \item The questionnaire included validation questions to confirm that respondents were developers~\citep{danilova2021_developers_questions} and possessed relevant software engineering experience.
    \item A tiered hourly compensation structure was implemented, with a \pounds 9 (“good”) remuneration for pre-screening and a \pounds 12 (“great”) remuneration for completing the main survey.\footnote{Prolific categorizes compensation into four tiers—low, fair, good, and great—based on the amount paid to participants.}
\end{itemize}

A cluster sampling strategy was employed to ensure that selected participants met predefined criteria aligned with the study's objectives. The target profile comprised (1) professionals employed in the Information Technology sector, (2) either full-time or part-time, (3) who had programming skills, and (4) held roles related to software engineering. To further ensure response quality and comprehension, respondents were required to (5) be fluent in English (the language of the survey) and (6) maintain a 100\% approval rate on Prolific with (7) at least 25 completed surveys, ensuring high reliability.\footnote{On Prolific, a participant’s approval rate reflects the percentage of accepted submissions, serving as a key indicator of reliability.}

The minimum required sample size was determined through an a priori power analysis using G*Power~\citep{faul_2009_GPower}. Assuming an effect size of 15\%, a significance level of 5\%, and a statistical power of 95\%, the required sample size for a model with eight predictors was calculated to be 160 participants. Since all participants from the previous study met these criteria, they were all invited to participate in this research. \revised{The first questionnaire, measuring the UTAUT2 constructs, was administered as part of our previous study~\citep{lambiase2024investigatingroleculturalvalues}. The second questionnaire, measuring the purpose-specific dependent variables (UB1--UB5), was administered between November 2024 and January 2025 by re-contacting the same participants through Prolific. The data collection was conducted in separate batches due to time constraints, and all 188 participants responded within the approximately three-month collection window.}

\subsubsection{Participants Characteristics}

The gender distribution of the 188 participants comprised 73.4\% men, 25.5\% women, and 1.1\% non-binary individuals. Participants were located across 13 countries, primarily from the United Kingdom (27.1\%), United States (26.6\%), Canada (10.1\%), Germany (9.6\%), and Spain (6.4\%), with additional representation from various European, Asian, African, and South American countries. The respondents' birth origins were primarily from North America (28.7\%), Western Europe (28.2\%), and Asia (17.0\%), reflecting a globally distributed sample.

Regarding professional roles, the most common positions were software developer/programmer (33.1\%) and software engineer (22.8\%), followed by data engineer (10.6\%), team leader (7.5\%), and project manager (6.6\%). Other roles, including tester/QA engineer (5.6\%), UX/UI designer (4.4\%), software architect (3.4\%), and others (1.6\%), contributed to the workforce diversity.

Experience levels varied across participants. In their current roles, most had 1-3 years (34.0\%) of experience, followed by 4-6 years (28.7\%) and more than 10 years (20.2\%). In the software industry overall, 34.0\% had over 10 years of experience, while 26.1\% had 1-3 years and 23.4\% had 4-6 years. This distribution highlights a mix of early-career professionals and seasoned experts.

\subsection{Data Analysis}

The data analysis procedure commenced with a \textit{preliminary evaluation} to ascertain data quality. This phase entailed assessing missing values, spotting unusual patterns, identifying outliers, and evaluating data distribution. Despite PLS-SEM's leniency compared to conventional techniques, these initial evaluations were crucial for confirming the data's reliability. No missing values and suspicious response patterns were detected. The distribution of the data was assessed for skewness and kurtosis to evaluate normality (values outside the range of -2 to +2); all values fell within acceptable limits, indicating no violations of normality assumptions.

Once validated, the dataset was imported into SmartPLS for PLS-SEM analysis~\citep{SmartPLS4}.\footnote{For a detailed explanation of the PLS-SEM process, refer to Hair et al.~\citep{hair_2014_PLS}.} The analysis followed a two-phase approach: \textit{Measurement Model Evaluation} and \textit{Structural Model Evaluation}.  

The \textit{Measurement Model Evaluation} phase assessed the reliability and validity of the constructs, ensuring that the items accurately reflected the underlying theoretical concepts. This step confirmed the adequacy of the measurement instruments before proceeding to hypothesis testing.  

The \textit{Structural Model Evaluation} phase examined the relationships between constructs, evaluating both the explanatory and predictive power of the model. Given that the model included mediated relationships—where a mediator transmits the effect of an independent variable (predictor) on a dependent variable (outcome)—both direct and indirect effects were analyzed. The primary focus was to determine which variables influencing developers' intention to adopt LLMs were associated with specific usage purposes.

\subsection{Threats to Validity}
\label{sec_limitation}

Here we present the threat to validity to our study.

\subsubsection{Conclusion Validity}

Conclusion validity pertains to the reliability of the relationships identified between independent and dependent variables~\citep{wohlin2012_experimentation}. The primary threats in this category arise from the statistical methods employed in the analysis. This study utilized PLS-SEM, a widely recognized method known for its robustness across diverse research contexts. Methodological rigor was ensured by strictly adhering to the guidelines outlined by Hair et al.~\citep{hair_2014_PLS}. Additionally, SmartPLS, a tool referenced in over \num{1000} peer-reviewed studies~\citep{hair_2014_PLS}, was used to enhance the reliability of the findings.

\subsubsection{Internal Validity}

Internal validity concerns the possibility that unaccounted-for factors influenced the dependent variables~\citep{wohlin2012_experimentation}. To mitigate this risk, the study was grounded in established and well-validated theories within the technology adoption domain. Moreover, strict participant selection criteria were applied to ensure that the sample accurately represented the target population while maintaining heterogeneity.

\subsubsection{Construct Validity}

Construct validity assesses the accuracy of the measurements~\citep{wohlin2012_experimentation}. All constructs were measured using validated instruments from prior research~\citep{Venkatesh_2012_UTAUT2}. The questionnaire was designed following state-of-the-art survey methodology, incorporating best practices such as question randomization and attention-check mechanisms to enhance reliability~\citep{kitchenham2008_PersonalOpinionSurveys, ralph_2020_empirical_standards, danilova2021_developers_questions}. Additionally, a well-established measurement instrument~\citep{eyal2021data, douglas2023data} was adopted, and a rigorous recruitment process was followed~\citep{alami2024you}.

\revised{A limitation of the measurement design concerns the use of single-item scales for the five dependent variables (UB1--UB5). While multi-item constructs are generally preferred in SEM studies to allow for reliability assessment, single-item measures can be appropriate when the construct is concrete and unambiguous---such as the self-reported frequency of a specific behavior~\citep{bergkvist2007measure, diamantopoulos2012guidelines}. This approach is consistent with the original UTAUT2 operationalization of Use Behavior by Venkatesh et al.~\citep{Venkatesh_2012_UTAUT2}, which we extended to purpose-specific usage. Additionally, single-item constructs are methodologically admissible in PLS-SEM. Nevertheless, we acknowledge that this choice limits the assessment of internal consistency for the dependent variables and may attenuate path coefficient estimates. Future studies should develop and validate multi-item scales for purpose-specific LLM usage to strengthen measurement reliability.}

\subsubsection{External Validity}

External validity pertains to the generalizability of the findings beyond the study sample~\citep{wohlin2012_experimentation}. To improve generalizability, rigorous filtering criteria were applied to the Prolific participant pool, ensuring that respondents met the necessary characteristics for this investigation. Furthermore, the required sample size was determined using G*Power recommendations~\citep{faul_2009_GPower} to ensure statistical adequacy. Nonetheless, generalizability remains a challenge, particularly given the evolving nature of generative AI technologies~\citep{Gartner2023_AI}. As the field matures, adoption patterns may shift, potentially leading to different findings in future research.

A threat concerns self-selection bias, as participation was voluntary. Despite applying strict filtering criteria to ensure respondents matched the target population, those who chose to participate may differ from those who opted out—for example in motivation or familiarity with LLMs. This may limit the representativeness of the results, although the recruitment protocol followed established, high-quality procedures to mitigate this issue.

Another threat relates to ecological validity. The study relies on self-reported survey data rather than observing LLM use in real development settings. As a result, responses may not perfectly capture how developers interact with LLMs during everyday workflows. While survey-based approaches are standard for large-scale SEM studies, future research could complement them with in-situ or log-based observations to strengthen ecological validity.

\revised{Moreover, this study focused exclusively on individual-level cognitive and behavioral factors as specified by UTAUT2. However, organizational policies, workplace rules, and team-level norms may also influence how developers adopt LLMs for specific purposes. For instance, restrictions on data sharing with external AI services, or mandated use of particular tools, could shape adoption patterns independently of individual attitudes. While our study partially mitigates this concern---by excluding participants whose LLM use was organizationally mandated, and through constructs such as Facilitating Conditions and Social Influence that capture aspects of the organizational context---a dedicated investigation of these factors remains warranted and represents an important direction for future research.}

\revised{Furthermore, the sample includes a mix of professional roles, including software developers, project managers, data engineers, and UX designers. While this heterogeneity reflects the multidisciplinary composition of modern software development teams and supports the generalizability of the findings, it also means that the analysis treats participants with potentially different patterns of LLM usage as a single group. Future studies could perform role-specific subgroup analyses to examine whether the influence of individual adoption factors varies across professional roles---for instance, whether social influence plays a different role for developers compared to project managers.}

\section{Analysis of the Results}

The following section presents the results of the PLS-SEM analysis. For readability and clarity, detailed information can be found in the online appendix \citep{online_appendix}, while the main findings are reported here.

\subsection{Measurement Model Evaluation}

As a first step in the evaluation of the theoretical model, it is paramount to evaluate the reliability of the constructs of the model~\citep{hair_2014_PLS, russo2021_pls_SLR}. 

\subsubsection{Indicator Reliability}
The evaluation of the measurement model begins with assessing indicator reliability by examining their \textit{outer loadings}. High outer loadings indicate that an indicator captures a substantial proportion of variance in the associated construct. A commonly accepted threshold is 0.708, as values above this level suggest that more than 50\% of the variance is shared with the construct. Indicators with loadings below 0.40 are generally removed, whereas those ranging between 0.40 and 0.70 are considered for removal if their exclusion enhances internal consistency reliability or convergent validity.

Only one indicator (FC\_4) did not meet the recommended threshold, displaying a loading of 0.665. However, since its value remained above 0.40 and its retention did not negatively impact internal consistency or validity, it was preserved for further analysis.

\begin{table}
    \centering
    \caption{Measurement Model Evaluation Metrics.} 
    \rowcolors{1}{white}{gray!15}
    \begin{tabularx}{\linewidth}{X c c c c}
    \toprule
        \textbf{Construct} & 
        \textbf{$\bm{\alpha}$} & 
        {$\bm{\rho_A}$} & 
        {$\bm{\rho_c}$} & 
        \textbf{AVE}
        \\ \midrule
        Performance Expectancy (PE) & 0.91 & 0.91 & 0.93 & 0.73 \\ 
        Effort Expectancy (EE) & 0.88 & 0.89 & 0.91 & 0.62 \\ 
        Social Influence (SI) & 0.82 & 0.83 & 0.87 & 0.58 \\ 
        Hedonic Motivation (HM) & 0.89 & 0.90 & 0.93 & 0.83 \\ 
        Habit (Hb) & 0.84 & 0.84 & 0.89 & 0.67 \\ 
        Facilitating Conditions (FC) & 0.75 & 0.75 & 0.84 & 0.57 \\
        Behavioral Intention (BI) & 0.90 & 0.90 & 0.94 & 0.83 \\ 
        
        \bottomrule
    \end{tabularx}
    
    \label{table_internal_consistency_reliability}
\end{table}

\subsubsection{Internal Consistency Reliability}
The second step focused on evaluating \textit{internal consistency reliability} to verify that the indicators consistently measure their corresponding constructs. This assessment was carried out using three key metrics: \textit{Cronbach's alpha} ($\alpha$), \textit{composite reliability} ($\rho_c$), and the \textit{reliability coefficient} ($\rho_A$).

As presented in Table \ref{table_internal_consistency_reliability}, all constructs surpassed the recommended threshold of 0.60 for these metrics~\citep{hair_2014_PLS}. These findings indicate that the indicators exhibit adequate internal consistency, ensuring a solid basis for subsequent analyses.

\subsubsection{Convergent Validity}  
Convergent validity assesses the degree to which indicators of a construct are positively correlated, indicating the extent of shared variance between them~\citep{hair_2014_PLS}. Since all constructs in the model were measured using reflective indicators, it was expected that the indicators would exhibit high convergence and substantial shared variance. The primary criterion for evaluating convergent validity is the \textit{average variance extracted} (AVE), where a value of 0.50 or higher suggests that the construct explains more than half of the variance in its associated indicators.

As presented in Table \ref{table_internal_consistency_reliability}, all constructs in the model attained AVE values exceeding the 0.50 threshold, confirming strong convergent validity and reinforcing the robustness of the measurement model.

\subsubsection{Discriminant Validity}
The final assessment examined \textit{discriminant validity}, which determines whether a construct is conceptually and empirically distinct from others. Henseler et al.~\citep{henseler2015_HTMT} proposed the \textit{heterotrait-monotrait ratio} (HTMT) as a robust criterion for evaluating discriminant validity. HTMT values are computed using the PLS-SEM algorithm, where values exceeding 0.90 indicate insufficient discriminant validity, values between 0.85 and 0.90 are considered acceptable, and values below 0.85 are regarded as excellent. Additionally, a \textit{bootstrapping} procedure is recommended to assess whether HTMT values significantly differ from the threshold values.

The detailed results of the PLS-SEM analysis are provided in the online appendix~\citep{online_appendix}. All HTMT values were below the recommended threshold of 0.85. Furthermore, a bootstrapping procedure was conducted with \num{10000} subsamples, using a one-tailed test and a significance level of 0.05. The results confirmed that all values remained below the critical thresholds, providing further evidence of sufficient discriminant validity.

\subsection{Structural Model Evaluation}

Following the evaluation of the measurement model, the structural model was assessed to examine the relationships between constructs and validate the hypotheses.

\subsubsection{Collinearity Analysis}
The initial step involved assessing collinearity between constructs to ensure reliable path estimations. To detect potential multicollinearity, the \textit{Variance Inflation Factor} (VIF) was used, a widely accepted measure in regression analysis. A VIF value below 3 is preferred, while values under 5 are typically considered acceptable.

In the analysis, most of the VIF values fell below the preferred threshold of 3, indicating minimal collinearity among variables. Only six values slightly exceeded this threshold, with the highest reaching 3.540, but all remained well within the acceptable upper limit of 5. These findings confirm that multicollinearity does not pose a significant issue in the structural model.

\begin{table*}
\centering
\caption{Direct Path Coefficients, Effect Sizes, and Significance—UB1--UB5}
\setlength{\tabcolsep}{3pt}
\rowcolors{1}{white}{gray!15}
\begin{tabularx}{\textwidth}{lXXXXXXXXXXXXXXX}
\toprule
\multirow{2}{*}{\textbf{Var}} 
    & \multicolumn{3}{c}{\textbf{UB1}} 
    & \multicolumn{3}{c}{\textbf{UB2}} 
    & \multicolumn{3}{c}{\textbf{UB3}} 
    & \multicolumn{3}{c}{\textbf{UB4}} 
    & \multicolumn{3}{c}{\textbf{UB5}} \\
\cmidrule(lr){2-4} \cmidrule(lr){5-7} \cmidrule(lr){8-10} \cmidrule(lr){11-13} \cmidrule(lr){14-16}
 & $\bm{\beta}$ & $\bm{f^2}$ & $\bm{p}$
 & $\bm{\beta}$ & $\bm{f^2}$ & $\bm{p}$
 & $\bm{\beta}$ & $\bm{f^2}$ & $\bm{p}$
 & $\bm{\beta}$ & $\bm{f^2}$ & $\bm{p}$
 & $\bm{\beta}$ & $\bm{f^2}$ & $\bm{p}$ \\
\midrule
PE & -0.070 & 0.003 & 0.514 & \textbf{-0.249} & \textbf{0.030} & \textbf{0.028} & \textbf{-0.273} & \textbf{0.032} & \textbf{0.016} & -0.145 & 0.009 & 0.248 & -0.098 & 0.004 & 0.415 \\
EE & -0.003 & 0.000 & 0.976 & \textbf{0.204} & \textbf{0.024} & \textbf{0.028} & \textbf{0.236} & \textbf{0.029} & \textbf{0.033} & \textbf{0.233} & \textbf{0.028} & \textbf{0.012} & 0.110 & 0.006 & 0.233 \\
SI & 0.096 & 0.012 & 0.164 & 0.142 & 0.023 & 0.070 & \textbf{0.253} & \textbf{0.068} & \textbf{0.001} & \textbf{0.187} & \textbf{0.036} & \textbf{0.014} & \textbf{0.159} & \textbf{0.024} & \textbf{0.049} \\
HM & 0.009 & 0.000 & 0.881 & 0.051 & 0.003 & 0.427 & 0.101 & 0.010 & 0.193 & 0.103 & 0.010 & 0.173 & 0.090 & 0.007 & 0.222 \\
Hb & \textbf{0.224} & \textbf{0.039} & \textbf{0.014} & \textbf{0.305} & \textbf{0.065} & \textbf{0.001} & 0.119 & 0.009 & 0.237 & 0.212 & 0.028 & 0.053 & \textbf{0.268} & \textbf{0.042} & \textbf{0.027} \\
FC & 0.092 & 0.009 & 0.164 & -0.084 & 0.007 & 0.267 & -0.188 & 0.032 & 0.054 & \textbf{-0.261} & \textbf{0.059} & \textbf{0.002} & \textbf{-0.221} & \textbf{0.040} & \textbf{0.009} \\
BI & -0.022 & 0.000 & 0.811 & -0.027 & 0.000 & 0.752 & 0.106 & 0.007 & 0.231 & 0.112 & 0.007 & 0.219 & -0.002 & 0.000 & 0.985 \\
UB & \textbf{0.501} & \textbf{0.290} & \textbf{0.000} & \textbf{0.410} & \textbf{0.177} & \textbf{0.000} & \textbf{0.353} & \textbf{0.119} & \textbf{0.000} & \textbf{0.201} & \textbf{0.037} & \textbf{0.027} & \textbf{0.271} & \textbf{0.064} & \textbf{0.015} \\
\bottomrule
\rowcolor{white}\multicolumn{16}{p{0.98\linewidth}}{\scriptsize Note: (1) Significant direct paths ($p<0.05$) are in bold. (2) $\beta$ is the standardized direct path coefficient; $f^2$ is Cohen's effect size (0.02 = small, 0.15 = medium, 0.35 = large); $p$ is the bootstrap p-value. (3) UB1 = ``Manipulate Artifacts'', UB2 = ``Generate Alternatives'', UB3 = ``Information Retrieval'', UB4 = ``Decision Support'', UB5 = ``Training''. (4) Values are direct effects only; for total (direct + indirect) effects, see Table~\ref{table_total_effect_significance_2}.} \\
\end{tabularx}
\label{table_direct_effects_f2}
\end{table*}

\begin{table}
\centering
\caption{Significance and Relevance of the Total Effect–BI and UB}

\begin{tabularx}{\linewidth}{lXXXX}
\toprule
\multirow{2}{*}{\textbf{Var}} 
    & \multicolumn{2}{c}{\textbf{BI}} 
    & \multicolumn{2}{c}{\textbf{UB}} \\
\cmidrule(lr){2-3} \cmidrule(lr){4-5}
{} & \textbf{T.E.} & $\bm{p}$ & \textbf{T.E.} & $\bm{p}$ \\
\midrule
\rowcolor{gray!15}PE & \textbf{0.443} & \textbf{0.000} & \textbf{0.116} & \textbf{0.003} \\
EE & 0.114 & 0.132 & 0.030 & 0.208 \\
\rowcolor{gray!15}SI & -0.061 & 0.261 & -0.016 & 0.283 \\
HM & 0.130 & \textbf{0.034} & 0.034 & 0.091 \\
\rowcolor{gray!15}Hb & \textbf{0.265} & \textbf{0.001} & \textbf{0.381} & \textbf{0.000} \\
FC & -0.015 & 0.803 & \textbf{0.158} & \textbf{0.008} \\
\rowcolor{gray!15}BI &  &  & \textbf{0.262} & \textbf{0.001} \\
\bottomrule
\rowcolor{white}\multicolumn{5}{p{0.95\linewidth}}{\scriptsize Note: (1) The significant relationships are in bold. (2) T.E. (Total Effect) refers to the combined product of all path coefficients connecting the construct to the dependent variable, both directly and indirectly.} \\
\end{tabularx}

\label{table_total_effect_significance_1}
\end{table}

\begin{table*}
\centering
\caption{Significance and Relevance of the Total Effect—UB1--UB5}
\rowcolors{1}{white}{gray!15}
\begin{tabularx}{\textwidth}{lXXXXXXXXXX}
\toprule
\multirow{2}{*}{\textbf{Var}} 
    & \multicolumn{2}{c}{\textbf{UB1}} 
    & \multicolumn{2}{c}{\textbf{UB2}} 
    & \multicolumn{2}{c}{\textbf{UB3}} 
    & \multicolumn{2}{c}{\textbf{UB4}} 
    & \multicolumn{2}{c}{\textbf{UB5}} \\
\cmidrule(lr){2-3} \cmidrule(lr){4-5} \cmidrule(lr){6-7} \cmidrule(lr){8-9} \cmidrule(lr){10-11}
{} & \textbf{T.E.} & $\bm{p}$ & \textbf{T.E.} & $\bm{p}$ & \textbf{T.E.} & $\bm{p}$ & \textbf{T.E.} & $\bm{p}$ & \textbf{T.E.} & $\bm{p}$ \\
\midrule
PE & -0.021 & 0.824 & \textbf{-0.213} & \textbf{0.045} & -0.185 & 0.071 & -0.073 & 0.515 & -0.067 & 0.526 \\
EE & 0.010 & 0.904 & \textbf{0.213} & \textbf{0.021} & \textbf{0.259} & \textbf{0.020} & \textbf{0.252} & \textbf{0.008} & 0.118 & 0.203 \\
SI & 0.089 & 0.189 & 0.137 & 0.080 & \textbf{0.241} & \textbf{0.002} & \textbf{0.177} & \textbf{0.020} & 0.155 & 0.054 \\
HM & 0.023 & 0.705 & 0.062 & 0.341 & 0.127 & 0.106 & 0.124 & 0.101 & 0.099 & 0.176 \\
Hb & \textbf{0.409} & \textbf{0.000} & \textbf{0.454} & \textbf{0.000} & \textbf{0.281} & \textbf{0.003} & \textbf{0.318} & \textbf{0.001} & \textbf{0.370} & \textbf{0.000} \\
FC & \textbf{0.172} & \textbf{0.015} & -0.019 & 0.798 & -0.133 & 0.188 & \textbf{-0.231} & \textbf{0.004} & \textbf{-0.179} & \textbf{0.030} \\
BI & 0.110 & 0.225 & 0.081 & 0.383 & \textbf{0.198} & \textbf{0.023} & 0.165 & 0.059 & 0.069 & 0.525 \\
UB & \textbf{0.501} & \textbf{0.000} & \textbf{0.410} & \textbf{0.000} & \textbf{0.353} & \textbf{0.000} & \textbf{0.201} & \textbf{0.027} & \textbf{0.271} & \textbf{0.015} \\
\bottomrule
\rowcolor{white}\multicolumn{11}{p{0.975\linewidth}}{\scriptsize Note: (1) The significant relationships are in bold. (2) T.E. (Total Effect) refers to the combined product of all path coefficients connecting the construct to the dependent variable, both directly and indirectly. Moreover, UB1 indicates “Manipulate Artifacts”, UB2 indicates “Generate Alternatives”, UB3 indicates “Information Retrieval”, UB4 indicates “Decision Support”, and UB5 indicates “Training”.} \\
\end{tabularx}

\label{table_total_effect_significance_2}
\end{table*}

\subsubsection{Significance and Relevance}

The structural model analysis assessed the significance and relevance of the hypothesized relationships. Significance testing was performed using the bootstrapping method with \num{10000} sub-samples, complete approach, and two tailed test type~\citep{hair_2014_PLS}. The analysis provided \textit{T-statistics}, \textit{p-values}, and significance levels for each hypothesized relationship.

Since the model included a series of mediated effects on the frequency of use for specific purposes, specifically mediated by \textit{intention to use} and the general \textit{frequency of use} of LLMs, the analysis was conducted in three consecutive steps:

\begin{enumerate} 
    \item First, the significance of the direct relationships between the UTAUT2 constructs and the frequency of use for specific purposes was assessed. 
    \item Next, the indirect relationships were examined, specifically those linking the constructs through either \textit{intention to use} (BI), \textit{frequency of use} (UB), or both. This approach allowed for the identification of whether the relationships between individual factors and the dependent variables were mediated, non-mediated, co-mediated, or competitively mediated. 
    \item Finally, the total effect of each factor on the dependent variables was analyzed to obtain a comprehensive measure that accounts for all the relationships. 
\end{enumerate}

For readability purposes, the results of this final analysis are presented in Table \ref{table_direct_effects_f2}, Table \ref{table_total_effect_significance_1} and Table \ref{table_total_effect_significance_2}. The following section outlines the findings by focusing on the dependent variables and explaining, for each, which independent variables proved to be significant and in what way.

\begin{itemize}[leftmargin=25pt]
    \item[UB1] Regarding the use of LLMs for generating or modifying artifacts, the relationships—ranked by relevance based on the product of path coefficients—indicate that UB exhibits a direct positive relationship, followed by Hb, which has a positive relationship through complementary mediation, and finally FC, which demonstrates a positive indirect relationship mediated through UB.

    \item[UB2] Regarding the use of LLMs for generating alternatives versions of an artifact, the ranking by relevance—determined using the product of path coefficients—shows that Hb has a positive relationship through complementary mediation, followed by UB with a direct positive relationship, then EE with a direct and unmediated positive relationship, and finally PE with a negative relationship due to competitive mediation through BI and UB. The presence of competitive mediation for PE, along with the strong effect of UB on UB2, suggests that the relationship between PE and UB2 is primarily explained through its mediated effect.

    \item[UB3] Moving to the use of LLMs for information retrieval, the ranking by relevance shows that UB has a direct positive relationship, followed by Hb and EE, both exhibiting positive relationships through complementary mediation via UB. Additionally, SI has a direct and unmediated positive relationship, while BI demonstrates a positive relationship through complementary mediation. Notably, UB3—along with UB4, as will be discussed shortly—represents the usage category with the highest number of significant relationships.

    \item[UB4] Moving to the use of LLMs for supporting decision-making, we observe that UB is displaced from its position as the most influential factor by Hb and EE, both of which exhibit direct positive, unmediated relationships. Next, we note the negative role of FC, which has a direct unmediated relationship. Following this, UB shows a positive relationship, and SI demonstrates a direct, unmediated positive relationship.

    \item[UB5] Finally, with regard to the use of LLMs for training purposes, Hb emerges as the central factor, showing a direct, unmediated positive relationship, followed by UB. Lastly, as observed previously, FC demonstrates a direct, unmediated negative relationship.
\end{itemize}

These findings reveal distinct patterns in how individual factors influence various uses of LLMs. While UB consistently plays a central role—an expected outcome given its representation of general usage frequency and so a symptom of statistical robustness—the diverse effects of Hb, EE, SI, PE, and FC highlight the complexity of adoption dynamics. Section \ref{sec_discussion} interprets these results, exploring their implications for research and practice.

\subsubsection{Explanatory Power}
In the final stage of the analysis, attention was paid to assessing the model's capacity to explain, specifically its effectiveness in fitting the data and assessing the strength of relationships within. The coefficient of determination, denoted R\textsuperscript{2}, served as the metric for evaluating the variance proportion accounted for by the model. R\textsuperscript{2} values span from 0 to 1, with higher values (above 0.19) indicating an increased explanatory power~\citep{chin1998_R_Square_value, raithel2012_explanatory_power_PLS_SEM, hair_2014_PLS}.

The analysis yielded R\textsuperscript{2} values of 0.463 for UB1, 0.409 for UB2, 0.350 for UB3, 0.329 for UB4, and 0.289 for UB5. These results indicate that the model provides a satisfactory level of explanatory power across all constructs, performing well for the first four use cases and comparatively less so for UB5. This suggests that while UTAUT2 effectively captures the key factors influencing the first four purposes, additional variables may be necessary to enhance the understanding of LLM adoption for training purposes.

\steSummaryBox{\faBarChart\ Summary of the Results.}{The analysis confirms that the model is both reliable and well-suited for predicting the dependent variables. As expected, UB plays a central role, while Habit emerges as a strong predictor. Furthermore, the varying significance of different factors across purposes reinforces the notion that adoption is not a uniform process, but rather a context-dependent phenomenon.
}


\begin{table*}
    \centering
    \caption{Summary of findings and implications for practice.} 
    \rowcolors{1}{white}{gray!15}
    \begin{tabularx}{\textwidth}{p{0.225\linewidth} XX}
    \toprule
    \textbf{Purpose} & \textbf{Findings} & \textbf{Implications}\\
    \midrule
    Generate Artifacts (UB1) & This purpose is minimally influenced by external factors. \textit{Habit} and \textit{facilitating conditions}—indirectly—play a role, suggesting it is a well-established, intuitive use case. & Given its widespread use, teams should focus on seamless LLM integration into workflows, such as IDEs and GenAI agents, to reinforce habitual adoption. \\
    Generate Alternatives (UB2) & \textit{Habit} and \textit{ease of use} are key drivers, with usability concerns playing a larger role than in artifact generation. Higher perceived difficulty discourages adoption. & Enhancing usability and accessibility of LLM-based modification tools can boost adoption. Improving interface design and automation provides a competitive advantage. \\
    Information Retrieval (UB3) & \textit{Social influence} directly drives adoption, making peer recommendations crucial. \textit{Habit} and \textit{ease of use} remain significant but secondary. & Organizations should foster peer-driven adoption through knowledge-sharing and validation. Encouraging cross-team adoption via \textit{boundary spanners} can accelerate LLM integration. \\
    Decision Making (UB4) & Unlike information retrieval, decision support depends directly on \textit{habit}, \textit{ease of use}, and \textit{social influence}. A negative relationship exists with perceived external support. & Decision-support LLMs should prioritize reliability and workflow integration. Organizations should use peer-driven strategies to enhance trust while balancing reliance on traditional resources. \\
    Training (UB5) & The lowest explanatory power suggests additional factors influence adoption. \textit{Habit} and \textit{usage frequency} matter, but perceived external support negatively impacts adoption. & Further research is needed, but promoting habitual engagement remains key. Organizations should explore structured AI-driven learning experiences to complement traditional training. \\
    \bottomrule
    \end{tabularx}

    \label{table_discussion}
\end{table*}

\section{Discussion and Implications}\label{sec_discussion}

In the following section, we discuss our results, providing interpretations, implications, limitations, and, also according to the last, future research directions.

\subsection{Advancements Concerning the Related Works}

Our work is based on previous contributions made in the context of research on LLM adoption for software development~\citep{agossah2023llm, draxler2023gender, russo2024_navigating, lambiase2024investigatingroleculturalvalues, choudhuri2024guideschoicesmodelingdevelopers} and aimed at corroborating previous results to advance the field of SE.

First, our work corroborates and extends the findings of Agossah et al.~\citep{agossah2023llm} and Draxler et al.~\citep{draxler2023gender}, who investigated general factors associated with LLM adoption. Agossah et al.~\citep{agossah2023llm} identified perceived usefulness, frequency of use, and social influence as key correlates of intention to use LLMs in a professional setting. Similarly, Draxler et al.~\citep{draxler2023gender} found that experienced users are more likely to adopt LLMs for professional purposes—such as generating formal content or retrieving technical information—and that habit and technological education play an important role in closing adoption gaps. Our study confirms and builds upon these findings by providing a purpose-specific analysis grounded in a structural model. Specifically, we show that perceived usefulness—captured as performance expectancy—is particularly influential in the adoption of LLMs for generating alternative versions of artifacts, while habit emerges as a robust predictor across all purposes. Furthermore, we refine previous insights on social influence by showing that its impact is not uniformly weak—as suggested by prior work—but becomes significant for purposes involving collaborative judgment, such as decision support.

Compared to our previous work~\citep{lambiase2024investigatingroleculturalvalues}, this study offers both corroboration and extension. In the earlier paper, we modeled LLM adoption as a general behavioral construct, placing individual cultural values at the center of the analysis and treating UTAUT2 constructs as mediators. We found that cultural values had no significant role, while performance expectancy, habit, facilitating conditions, and behavioral intention emerged as key drivers of adoption. The current study confirms these findings—thus reinforcing their validity—but also advances them by showing how these factors vary depending on the specific purpose for which LLMs are used. This purpose-specific perspective reveals that the influence of individual factors is not homogeneous: their impact depends on the developer’s goals and context of use. These insights offer more actionable guidance for organizations, allowing them to tailor their interventions based on the intended use of LLMs.

Our study also builds on and refines the findings of Russo~\citep{russo2024_navigating} and Choudhuri et al.~\citep{choudhuri2024guideschoicesmodelingdevelopers}. Russo showed that traditional predictors from models like TAM and UTAUT—such as perceived usefulness and social influence—play a marginal role in LLM adoption, while compatibility with existing workflows emerges as the key driver~\citep{russo2024_navigating}. We confirm and extend this insight by demonstrating that habit and facilitating conditions are the most consistent predictors across purposes, emphasizing the importance of seamless integration into daily work. In parallel, Choudhuri et al.~\citep{choudhuri2024guideschoicesmodelingdevelopers} found that trust in generative AI systems—driven by output quality, goal alignment, and functional value—is the strongest determinant of adoption. While our model does not include trust explicitly, we show that for cognitively demanding purposes such as training and decision support, adoption is driven by performance expectancy, habit, and facilitating conditions—factors that closely reflect the conditions under which trust is likely to form. Crucially, our purpose-specific analysis advances both prior works by revealing that the weight of each predictor is not uniform: for instance, social influence—dismissed in previous studies—proves significant for decision-oriented purposes. Our model also articulates a full causal pathway from individual perceptions to usage behavior, highlighting how general intentions translate into concrete, purpose-level adoption.

\revised{Our work also complements the work of Brown et al.~\citep{brown2024trust}; while their contribution provides valuable fine-grained behavioral insights at scale, our study advances this line of work in three complementary ways: (1) we examine adoption across five distinct purposes of LLM use rather than a single behavior; (2) we investigate the cognitive and perceptual antecedents of adoption using a validated theoretical framework rather than behavioral logs alone; and (3) our structural equation modeling approach enables the analysis of direct, mediated, and total effects, offering insight into how individual factors translate into purpose-specific usage.}

Overall, our findings suggest that LLM adoption is complex not only in terms of the factors that influence it (e.g., performance expectancy, habit, or social influence), but also in terms of what it actually means to adopt. \faHandORight\ \textbf{Organizations that perceive weaknesses in specific areas—and expect LLMs to support them primarily in those—can use our findings and practical implications as targeted guidance to improve efficiency and effectiveness}. At the same time, \faBook\ \textbf{researchers investigating LLM adoption can leverage our results to better design empirical studies and to explain heterogeneous usage patterns.}

\subsection{Interpretation of Findings and Implications}

The findings provide a nuanced understanding of how individual factors influence the adoption of LLMs for specific software engineering purposes. These are outlined in the following and summarized in Table~\ref{table_discussion}.

\subsubsection{\underline{On the Central Role of Frequency of Use}}
As an initial consideration, the results confirm the centrality of \textit{general usage frequency} (\textit{UB}) across all purposes. As previously mentioned, this result is not surprising; rather, the opposite would have raised a significant conceptual inconsistency—beyond a theoretical one—in the research model. Indeed, it is entirely reasonable that the overall frequency of LLM usage within the software development lifecycle is positively associated with its frequency of use for specific purposes. Conceptually, this result is straightforward to justify and serves primarily as a validation of the model's robustness and the reliability—already extensively discussed in the methodology section (Section \ref{sec:experiment})—of the data collection and analysis process.

From a practical perspective, it is evident that \faHandORight\ \textbf{an overall increase in the frequency of LLM usage, and thus a greater number of opportunities to engage with these tools, can naturally lead to a higher frequency of use for specific purposes}. This dynamic is often observable in practices such as prompting LLMs during routine activities like writing Git commit messages, updating documentation, or generating boilerplate code—activities that, due to their low risk and high frequency, naturally reinforce habitual use across the development workflow.

Moving away from the practical implication, the most compelling aspect of this result is that the strength and mediating role of \textit{UB} are not uniform across all purposes. This variation underscores the influence of additional factors—both organizational and cognitive—that shape the adoption of LLMs for specific SE purposes, which will be examined in more detail in the following sections.

\subsubsection{\underline{Using LLMs to Manipulate Artifacts}}
Regarding the use of LLMs to generate or modify artifacts (UB1), this purpose emerges as one of the least influenced by the various adoption drivers. Apart from the frequency of general use, only \textit{habit} and \textit{facilitating conditions}—albeit indirectly—prove to be significant factors. To interpret this result, it is important to emphasize that artifact generation, such as source code or documentation, is one of the most established use cases for these tools and is widely perceived as enhancing productivity~\citep{liu2024empirical, Jiang2023Self_planning}. This suggests that the use of LLMs for this purpose is so generic that it is not strongly dependent on specific adoption factors but rather emerges as an activity that naturally integrates into software development workflows. 


\revised{\faHandORight\ \textbf{Nevertheless, teams exploring the integration of LLMs for this broad purpose can benefit from embedding them into routine development environments in a way that supports developers' existing workflows.} Integrating LLMs into everyday activities---such as through IDE plug-ins, in-line suggestions during code authoring, or integration into pull request automation---can foster habitual use. For tool designers and tech leads, aligning LLM-based features with common artifact manipulation activities (e.g., code refactoring, docstring generation) can help ensure that adoption is frictionless and perceived as productivity-enhancing rather than disruptive. This is particularly relevant in routine scenarios such as adjusting configuration files, updating boilerplate components, or performing minor edits during code review---situations where LLMs can assist with repetitive changes while allowing developers to remain in control and to assess the quality of generated outputs before accepting them.}

\subsubsection{\underline{Using LLMs to Modify Artifacts}}

Regarding the use of LLMs for generating alternatives of a given artifacts, \textit{habitual use} and \textit{perceived ease of use} emerge as the primary drivers of adoption. This pattern closely mirrors the findings for artifact generation, which is expected given the conceptual similarity between the two purposes. However, a key distinction is that while habitual use is the dominant factor in artifact generation, in this case, where a greater conceptual effort is required—\textit{perceived effort} begins to play a role. Engineers who perceive LLMs to be difficult to use are likely to prefer to generate alternatives manually.


\revised{\faHandORight\ \textbf{A practical implication of this finding is that LLM-based tools designed for generating alternatives should prioritize ease of use.} This is especially important in tasks that require conceptual reinterpretation or creative variation (e.g., code restructuring, UI redesign, documentation variants), where perceived friction may discourage engagement. Ensuring low-effort access to high-quality suggestions within the IDE---combined with intuitive interaction patterns (e.g., accepting, editing, or rejecting with minimal steps)---can improve perceived usability and lower the entry barrier for developers.} 

\faHandORight\ \textbf{Similarly, GenAI tool developers should treat usability as a competitive differentiator.} Seamless integration, contextual suggestions, and responsiveness to user intent are essential to support habitual engagement with LLMs across artifact modification tasks. For example, during requirements elicitation or refinement, offering context-aware completions for user stories or acceptance criteria can make these tools more useful and aligned with early-phase development workflows.

\subsubsection{\underline{Using LLMs to Retrieve Information}}

For retrieving information (\textit{UB3}), the results present a more intriguing picture. Given that this purpose requires trust in the tool, habitual use, and perceived ease of use remain dominant predictors, as observed in the previous cases. However, a particularly noteworthy finding is the role of \textit{social influence}, which not only exerts a strong effect but does so in a direct and unmediated manner, meaning its impact is entirely attributable to the construct itself. This result suggests that the adoption of LLMs for information retrieval is not solely an individual decision driven by usability and familiarity but is also strongly shaped by external influences within the professional environment. Unlike artifact generation or modification—where efficiency and personal workflow integration may dominate the decision to use LLMs—information retrieval inherently involves elements of credibility. The strong, direct impact of \textit{social influence} suggests that professionals are more likely to adopt LLMs for retrieving information when they observe or receive endorsements from peers, team leaders, or authoritative figures within their organization. This aligns with existing research on technology adoption in knowledge-intensive tasks, where peer recommendations and organizational norms play a crucial role in shaping behavior~\citep{Eckhardt2009Who}.


\revised{\faHandORight\ \textbf{For information retrieval, our findings suggest that tool designers and team leaders should consider fostering environments where usage is visible, shared, and socially reinforced.}} In knowledge-sensitive tasks—such as architectural decisions, bug triage, or onboarding—developers often rely on peer validation to trust retrieved information. Endorsement by senior developers, internal champions, or influential peers can act as catalysts by legitimizing use and providing situated examples. \revised{\faHandORight\ \textbf{Additionally, organizations may benefit from leveraging boundary spanners---developers who naturally operate across teams and share practices.}} These individuals can play a strategic role in spreading effective LLM usage patterns, especially in distributed or cross-functional environments. For example, a backend engineer contributing to multiple microservices teams might introduce prompt engineering strategies or tool configurations that proved effective in one context, helping standardize LLM-assisted information retrieval across the organization.

It is also important to recognize the potential impact of LLM hallucinations in this context. Because information retrieval inherently depends on perceived credibility, hallucinated or misleading outputs can disrupt both individual trust and socially mediated validation mechanisms. For example, if a developer retrieves incorrect API usage guidance from an LLM and this is noticed during code review or team discussion, the perceived reliability of the tool—and the developer’s willingness to rely on it—may decrease. \faBook\ \textbf{Future work should investigate how negative experiences with hallucinated outputs affect sustained adoption patterns, especially in socially influenced settings.}

\subsubsection{\underline{Using LLMs to Support Decision Making}}

Regarding the use of LLMs for supporting decision-making (\textit{UB4}), the results initially appear similar to those observed for information retrieval (\textit{UB3}), with \textit{habit}, \textit{perceived ease of use}, and \textit{social influence} emerging as key adoption drivers. However, a crucial difference lies in the nature of these relationships. While in the previous case, \textit{habit} and \textit{perceived ease of use} exhibited a \textit{complementary mediated} relationship—meaning their impact was reinforced through the central and consistent role of general usage frequency (\textit{UB})—in the context of decision support, these factors, along with \textit{social influence}, exert a \textit{direct and unmediated} effect. This central difference likely stems from the fundamental nature of these purposes and the cognitive processes they involve. While in information retrieval, adoption is reinforced through habitual use and frequent exposure, fostering a gradual trust-building process, in decision support, reliance on LLMs is more immediately shaped by perceived ease of use and social validation, as users face higher cognitive demands and potential risks, making trust a prerequisite rather than an outcome of repeated engagement.


These results have direct implications: the potential central role of \textit{reliability}—coherent with the work of Choudhuri et al.~\citep{choudhuri2024guideschoicesmodelingdevelopers}.
\faHandORight\ \textbf{To support LLM adoption in decision-making, developers need tools that are both reliable and deeply integrated into their daily workflows.} Given the cognitive demands and perceived risk associated with delegating choices to AI, usability alone is insufficient—perceived trustworthiness must be central to design. For example, surfacing confidence scores, showing source attribution, or enabling human-in-the-loop validation can help developers better assess the reliability of LLM suggestions in tasks such as architecture decisions or code reviews.

Additionally, the direct impact of \textit{social influence} suggests that users seek validation from their peers when incorporating LLMs into decision-making. \revised{\faHandORight\ \textbf{Organizations may benefit from peer-driven strategies to support informed adoption, such as internal testimonials, expert endorsements, or shared use-case libraries.}} Embedding LLMs into existing decision-support practices—e.g., sprint planning or RFC review sessions—can normalize their use and reduce resistance rooted in uncertainty or organizational culture.

Moreover, a direct, and more intriguing, negative relationship emerged between decision support adoption and the \textit{perception of working in an environment that facilitates general LLM adoption}. This result is particularly interesting and somewhat counterintuitive. Naturally, the intention is not to suggest that reducing available resources would lead to greater adoption—an argument that would be unethical and irrational. Instead, a plausible explanation is that in environments where engineers heavily rely on external resources and perceive a highly supportive and facilitating context, they may prefer to use these existing resources for decision-making rather than placing trust in an AI-driven tool. This further reinforces the central role of \textit{reliance} in LLM adoption for specific purposes, suggesting that when alternative sources of decision support are readily available, engineers may be less inclined to depend on LLM-generated recommendations.

Furthermore, because developers are using LLMs to support cognitively demanding and high-impact decisions, inaccurate or fabricated outputs can severely undermine trust and hinder sustained adoption. This aligns with broader concerns in the literature regarding the limitations of current LLMs in providing verifiable and context-aware suggestions. \faBook\ \textbf{Future research should examine how different reliability-enhancing mechanisms—such as explanation interfaces or feedback loops—can mitigate the effect of hallucinations and promote informed, rather than blind, reliance.}

\subsubsection{\underline{Using LLMs to Train and Learn}}

Regarding the training purpose (\textit{UB5}), it is important to highlight its role as the least well-defined purpose, as indicated by the lowest \textit{R\textsuperscript{2}} value among all dependent variables. Beyond the consistently significant influence of \textit{habit} and \textit{general usage frequency}, a negative relationship also emerged between training adoption and the perception of available support resources. As previously discussed, this suggests that in environments where professionals perceive strong external support, they may rely more on traditional learning resources rather than integrating LLMs into their training routines.

However, it is crucial to acknowledge that additional factors, not captured by our model also for model parsimony reasons~\citep{hair_2014_PLS}, may influence the adoption of LLMs for training purposes. While the findings suggest that promoting habitual use remains a relevant strategy, further research is necessary to better understand the specific drivers of LLM adoption in learning and professional development contexts.

\subsection{Gaps and Future Research Directions}

As with any research, several limitations must be acknowledged, representing opportunities for future research.

First, the relatively lower \textit{R\textsuperscript{2}} value for \textit{UB5 (Training)} suggests that additional factors, such as \textit{learning self-efficacy}, \textit{cognitive load}, or \textit{motivation to learn}, may be necessary to improve explanatory power. \faBook\ \textbf{Future studies could integrate these constructs to refine the understanding of LLM adoption in learning and professional development contexts}, particularly in environments where alternative training resources are readily available.

Second, while UTAUT2 effectively captures broad behavioral trends, the results indicate that LLM adoption is not uniform across purposes, supporting the argument that adoption is a domain-specific phenomenon rather than a generalizable trend. \faBook\ \textbf{Consequently, incorporating purpose-specific adoption factors—such as \textit{error tolerance}, \textit{perceived AI reliability}, or \textit{trust in automation}—could improve the explanatory and predictive power of future models.} Research focusing on how different SE subdomains (e.g., testing, debugging, requirements engineering) shape LLM adoption would further enhance theoretical and practical insights.

Expanding on the previous point, the finding also reflects limitations inherent to the UTAUT2 framework, which—like other behavioral models rooted in the Technology Acceptance Model (TAM)—tends to emphasize prediction over contextual explanation and may not fully capture the situated and interpretive dimensions of LLM use. Such models are grounded in a positivist tradition that seeks to explain behavior through predefined, generalizable constructs, often overlooking the complexity of technology-in-practice. These concerns have been raised in the information systems literature, particularly in calls for theoretical perspectives that make the role of technology in practice more explicit and entangled with organizational context~\citep{orlikowski2001desperately, orlikowski2007sociomaterial}. \faBook\ \textbf{We therefore see our work as a foundational contribution, and encourage future studies to build on this foundation by adopting interpretivist or sociomaterial approaches to better capture how LLMs are situated and appropriated in actual SE contexts.}

Third, the model does not account for the quality of LLM outputs, particularly the risk of hallucinations—i.e., plausible but inaccurate or fabricated responses~\citep{Tonmoy2024A, Liu2024Exploring}. As LLMs are increasingly used in tasks that rely on credible information or support high-stakes decision-making, these inaccuracies can erode trust and disrupt adoption dynamics. \faBook\ \textbf{Future studies could explore how developers perceive, detect, and manage hallucinated outputs, and how these experiences affect trust, reliance, and long-term use of LLMs in specific software engineering contexts.} Integrating constructs such as perceived output reliability, confidence calibration, or prior negative experiences could offer a more nuanced view of adoption drivers.

Fourth, this study relied on self-reported survey data, which, while widely used in technology adoption research, may introduce biases related to self-perception, social desirability, and recall accuracy. \faBook\ \textbf{Future work could complement survey data with behavioral usage metrics, such as interaction logs or real-time feedback from LLM-based development environments.} Combining qualitative approaches, such as longitudinal observations, with quantitative methods could provide a more comprehensive view of adoption dynamics.

\revised{Additionally, the study does not explicitly account for organizational-level factors---such as company policies on AI tool usage, managerial directives, or team-level norms---that may independently influence LLM adoption for specific purposes. While participants whose LLM use was mandated were excluded, and constructs such as Facilitating Conditions and Social Influence partially capture organizational context, these proxies may not fully represent the range of organizational influences at play.}

Furthermore, the strong role of social influence in the adoption of information retrieval and decision-making raises the need for further investigation into the role of team dynamics. \faBook\ \textbf{Future research could explore how factors such as team size, hierarchy, and collaboration patterns impact LLM adoption within SE teams, particularly in distributed or cross-functional settings.}

Finally, given the evolving nature of generative AI, adoption patterns may shift as technology matures. \faBook\ \textbf{Future studies should track longitudinal adoption trends, examining how increasing familiarity, improvements in model accuracy, and evolving industry practices influence LLM adoption over time.} Research in this area could inform best practices for integrating LLMs into SE workflows while addressing emerging challenges such as AI-driven decision-making, responsibility attribution, and ethical considerations.

\section{Conclusion}

The findings of this study underscore that LLM adoption for SE purposes is not uniform across different activities, reinforcing the argument that adoption is a domain-specific phenomenon rather than a broadly generalizable trend. While \textit{habitual use} consistently emerges as a key driver, different purposes are influenced by distinct adoption factors. For instance, the mediation effects observed in performance-driven purposes (UB1, UB2) suggest that LLM adoption follows an iterative, reinforcement-driven trajectory, where continued engagement strengthens its perceived usefulness over time.
Conversely, the significant role of \textit{facilitating conditions} in purposes centered on information retrieval (UB3) and processing (UB4) highlights the complexity of LLM adoption in SE. This suggests that beyond habitual engagement, structural and organizational factors shape whether professionals integrate LLMs into their workflows.

Additionally, the results align with and complement prior studies~\citep{russo2024_navigating, lambiase2024investigatingroleculturalvalues, choudhuri2024guideschoicesmodelingdevelopers}, particularly the work of Khojah et al.~\citep{khojah2024beyond}. While their theoretical framework traces adoption from purpose of adoption to personal experience, the present findings extend this perspective by identifying the antecedents that drive purpose-specific adoption in the first place. This positioning broadens the conceptual understanding of LLM adoption in both research and practice, offering a more comprehensive view of this disruptive phenomenon.

Finally, in addition to outlining future research directions to address the study’s limitations, all data used in this research, including the materials employed for data collection, are publicly available (through an online appendix). This ensures replicability, facilitates further advances, and enhances the reliability of the findings.

\section*{Acknowledgment}

The authors thank the questionnaire participants for their valuable responses and for providing the essential data required to conduct this research. We acknowledge the use of ChatGPT-4 to ensure linguistic accuracy and enhance the readability of this article.

\bibliographystyle{elsarticle-num-names}
\bibliography{reference}

\end{document}